\documentclass[reprint,aps,floatfix,nofootinbib]{revtex4-2}

\usepackage{amsbsy}
\usepackage{amssymb}
\usepackage[sumlimits]{amsmath}
\usepackage[unicode]{hyperref}
\usepackage{cleveref}
\usepackage{graphicx}
\usepackage[active]{srcltx}
\usepackage{pdfsync}
\usepackage{float}

\usepackage{color}

\begin{document}

\title{Detecting the impact of nuclear reactions on neutron star mergers through gravitational waves}

\author{P. Hammond, I. Hawke and N. Andersson}

\affiliation{
Mathematical Sciences and STAG Research Centre, University of Southampton,
Southampton SO17 1BJ, United Kingdom}

\begin{abstract}
Nuclear reactions may affect gravitational-wave signals from neutron-star mergers, but the impact is uncertain. In order to quantify the effect, we compare two numerical simulations representing intuitive extremes. In one case reactions happen instantaneously. In the other case, they occur on timescales much slower than the evolutionary timescale. We show that, while the differences in the two gravitational-wave signals are small, they should be detectable by third-generation observatories. To avoid systematic errors in equation of state parameters inferred from observed signals, we need to  accurately implement nuclear reactions in future simulations.
\end{abstract}

\maketitle

\section{Introduction}

Binary neutron star mergers are promising cosmic laboratories for extreme physics. As demonstrated in the celebrated case of GW170817 \cite{Abbott2017}, one can use observed gravitational-wave merger signals (along with any electromagnetic counterparts) to make progress on the vexing issue of the state of matter at densities beyond nuclear. This motivates much of the current effort to develop robust numerical simulation technology (in full general relativity) to model these events \cite{Baiotti2017,Bernuzzi2020}, including as much realistic physics as possible. Reliable numerical relativity (and general relativistic magnetohydrodynamics) simulations are required to model the highly non-linear dynamics at play and produce the signal templates needed for parameter extraction. 
In parallel, we need to improve the detector technology. The gravitational-wave signal from a merger event is characterised by frequencies above $1\ \mathrm{kHz}$, where current ground-based instruments lose sensitivity. 

While current interferometers could possibly detect the post-merger gravitational-wave signal from a nearby event, one would have to be very lucky for such an event to take place and with the spectacular GW170817 observation we may have been as fortunate as we are going to get, and in that case we did not observe the post-merger dynamics. In order to make our own luck, we need to push the development of third-generation instruments (like the Einstein Telescope \cite{Punturo2010,Hild2011,ET} and the Cosmic Explorer \cite{Reitze2019,CE}) which will have the potential to regularly catch neutron-star mergers \cite{Baiotti2022}. 

At the same time, we need to make progress on the issue of extracting the physics we want to explore from observations. Motivated by this, simulations of large sets of mergers involving different matter equations of state (and physics implementations) have been carried out \cite{Dietrich2018}. The results demonstrate how one may, indeed, expect to be able to distinguish different matter descriptions. This is promising, but a few points of caution are in order. In particular, we need to keep in mind that current simulations are not yet able to represent all aspects of the expected reality. Given the level of difficulty of the issues involved (especially concerning neutrinos \cite{Cabezn2018,Pan2018,Richers2017,Cusinato2021,Mezzacappa2020}) progress is slow. 
As we improve our simulations, we need to ensure that the results are sufficiently accurate that the signal templates do not introduce systematic errors due to un- or under-resolved physics. This raises questions somewhat different from that of comparing different matter equations of state. We need to ask 
if different physics implementations \emph{for the same equation of state} can be observationally distinguished. This question motivates the present work. 

In this letter we focus on the impact of nuclear reactions on the gravitational wave-signal recovered from simulations. In the simplest case, the reactions are responsible for the balance between neutrons, protons, and electrons in the high-density matter, and are also responsible for production of neutrinos in neutron stars (which in turn leads to bulk viscosity \cite{Celora2022}). While there are current efforts \cite{Radice2022,Camilletti2022,Palenzuela2022,Hayashi2021,Cusinato2021} to include neutrinos in neutron star merger simulations,  we are only aiming to get a handle on the ``error bars'' involved so we take a more simplistic approach. We consider two ``limiting cases'', such that the reactions either take place on such a long timescale that the rates can be set to zero, or on such a short timescale that the fluid composition reaches equilibrium instantaneously. Our results demonstrate that gravitational-wave signals from these two simulations---although very similar---should be distinguishable with a third-generation instrument like the Einstein Telescope. The implications are simple. We either have to accept the difference as a systematic error or we need to make progress on implementing nuclear reactions in our simulations.

\section{Method}

\textit{Simulating reaction limits}.---
We assume the neutron star fluid is composed of neutrons, protons, and electrons, which have number densities $n_\mathrm{n}$, $n_\mathrm{p}$, and $n_\mathrm{e}$ respectively, and that the fluid is locally charge neutral, which imposes the condition $n_\mathrm{p} = n_\mathrm{e}$. We then define the baryon number density as $n_b = n_\mathrm{n} + n_\mathrm{p}$, proportional to the fluid rest mass density $\rho$. To keep track of the composition of the fluid we use the electron fraction $Y_\mathrm{e}$ defined by
\begin{align}
    Y_\mathrm{e} = \frac{n_\mathrm{e}}{n_b} = \frac{n_\mathrm{p}}{n_b}.
\end{align}
The standard Valencia formulation~\cite{Banyuls1997, Baiotti2005} of the general relativistic evolution equations already accounts for the evolution of $n_b$ through $\rho$, however an extra equation is required for the evolution of $Y_\mathrm{e}$. This takes the form
\begin{align}
    u^a \nabla_a Y_\mathrm{e} &= \frac{\Gamma_\mathrm{e}}{n_b}, \label{eqn:ye_evolution_full}
\end{align}
where $u^a$ is the fluid four velocity, and $\Gamma_\mathrm{e}$ is the total rate of electron production (for the simplest case of the Urca processes this will be the rate of neutron decay minus the rate of electron capture). The left hand side of the equation accounts for the advection of the composition along the world line of a given fluid element, and the right hand side accounts for the change in this composition due to reactions.

While this relation is simple to write down, the interaction between the rates and the numerical methods make it much less simple to evolve. If the reaction rate $\Gamma_\mathrm{e}$ is large compared to $n_\mathrm{e}\Delta t $ (where $\Delta t$ is the simulation timestep), or equivalently the equilibriation timescale of the reactions is fast compared to $\Delta t$, then \cref{eqn:ye_evolution_full} becomes stiff and explicit numerical evolution schemes produce unphysical oscillations in $Y_\mathrm{e}$. The standard work-arounds for this (decrease $\Delta t$, increase convergence order, use an implicit scheme, etc.) are either computationally expensive or would require significant changes to the method used (or both), so as a first step we instead work in the two limits of $\Gamma_\mathrm{e}$: the fast reaction limit where $\Gamma_\mathrm{e} \rightarrow \infty$, and the slow reaction limit where $\Gamma_\mathrm{e} \rightarrow 0$.

In the slow reaction limit \cref{eqn:ye_evolution_full} simplifies to a standard advection equation where fluid composition is preserved along the world lines of fluid elements, while the assumption that the reactions occur on timescales too fast to be resolved by the simulation will cause the matter to always be in $\beta$-equilibrium. The definition of this equilibrium (in terms of chemical potentials) changes as the temperature of the fluid varies due to different reactions dominating the changes in composition~\cite{Alford2018a,Hammond2021}. As we are not yet able to implement this effect---we do not have access to the extra chemical potentials required for the other equilibria \cite{Alford2018a}---we bypass the issue by using (for the fast-limit simulation)  the ``cold'' $\beta$-equilibrium, $\mu_\mathrm{n} = \mu_\mathrm{p} + \mu_\mathrm{e}$.

We describe our current simulation setup in detail in~\cite{Hammond2021}. Briefly, we use a standard set of thorns within the Einstein Toolkit~\cite{EinsteinToolkit}  modified to facilitate the use of 3-parameter tabulated equations of state. The simulation uses adaptive mesh refinement for the grid, with the highest spacial resolutions grids being centered on and completely covering each star with a grid spacing of $\sim 400\,\mathrm{m}$, and the timestep we use is $\Delta t / \Delta x=0.25$. A single set of initial data for two stars, each with baryon mass $M_\mathrm{b} = 1.4\,M_\odot$, separated by $40~\mathrm{km}$ obtained using LORENE~\cite{LORENE} is evolved in all simulations presented here, assuming a fixed initial temperature of $T=0.02\,\mathrm{MeV}$ and cold $\beta$-equilibrium. No magnetic fields are present. 

In order to keep the simulations as comparable as possible, we use identical code for the in- and out-of-equilibrium simulations, and account for the instantaneous equilibriation through the equation of state table. In order to achieve this we take the 2D slice defined by $\mu_\mathrm{n} = \mu_\mathrm{p} + \mu_\mathrm{e}$ through the full 3D table, then replace each value of $p(\rho,T,Y_\mathrm{e})$ in the 3D table with the value $p(\rho,T)$ found on the 2D slice (and equivalently for other equation of state variables). We can then use this table with the same interface as the original table. 

\textit{Gravitational wave analysis}.---
We measure the gravitational radiation from the merger through the Weyl scalar $\Psi_4 = \ddot{h}_{+} - i \ddot{h}_\times$, using the \textit{Fixed Frequency Integration} method of Reisswig and Pollney~\cite{Reisswig2011} to obtain the raw strain $h = {h}_{+} - i {h}_\times$. As the stars remain close to equilibrium until merger, we do not expect any significant differences to appear during the inspiral (and indeed we do not see significant differences between the simulations in the $\sim 5$ pre-merger orbits covered), hence we focus our analysis on the post-merger signal. We do this in two ways: in one method we ignore any data before the maximum $\left|h\left(t\right)\right|$, in the second we apply a high-pass filter to all of the data.

To quantify the impact of the reaction limits on the recovered signals we use a standard measure of the difference between two gravitational-wave signals, the mismatch $\mathcal{M}$ (see e.g.~\cite{Lindblom2008,McWilliams2010,Baird2013,Kumar2015}). The mismatch is calculated through the waveform overlap, $\left\langle h_1|h_2\right\rangle$, defined by
\begin{align}
    \left\langle h_1|h_2\right\rangle &= 4 \int^{\infty}_{-\infty} \frac{\tilde{h}_1\left(f\right) \tilde{h}^{\ast}_2\left(f\right)}{S_n \left(f\right)} \mathrm{d} f, \label{eqn:overlap}
\end{align}
where $\hat{h}\left(f\right)$ is the Fourier transform of $h$ and $\ast$ denotes its complex conjugate, while $S_n$ is the strain sensitivity of a detector as power spectral density. As the simulation outputs a discretely sampled signal, we discretise \cref{eqn:overlap} to obtain
\begin{align}
    \left\langle h_1|h_2\right\rangle &= 4 \Delta f \sum^{f_\mathrm{max}}_{f=f_\mathrm{min}}\left( \frac{\tilde{h}_1\left(f\right) \tilde{h}^{\ast}_2\left(f\right)}{S_n \left(f\right)}\right), \label{eqn:overlap_discrete}
\end{align}
where the tilde denotes the discrete Fourier transform (normalised for consistency with the continuous version), $f_\mathrm{max}$ is the maximum resolvable frequency in the data (typically the Nyquist frequency), $f_\mathrm{min}$ is the minimum resolvable frequency (which for us is either the $\omega_\ast$ used in the fixed frequency integration, or the filter frequency), and $\Delta f$ is the frequency resolution of the discrete Fourier transform.

The match $\mathrm{M}$ (sometimes referred to as the ``maximised overlap'') between the two signals is given by
\begin{align}
    \mathrm{M} &= \frac{\max\left(\left\langle h_1|h_2\right\rangle\right)}{\sqrt{\left\langle h_1|h_1\right\rangle \left\langle h_2|h_2\right\rangle}}, \label{eqn:match}
\end{align}
where the overlap between $h_1$ and $h_2$ should be maximised by time and phase shifting $h_2$ by $(t_c, \phi_c)$. The mismatch $\mathcal{M}$ is then given by one minus the real part of the (maximised) match, so we obtain
\begin{align}
    \mathcal{M} &= 1 - \Re \left(\frac{\max\left(\left\langle h_1|h_2 e^{i\left(\phi_c-2 \pi f t_c\right)} \right\rangle\right)}{\sqrt{\left\langle h_1|h_1\right\rangle \left\langle h_2|h_2\right\rangle}}\right). \label{eqn:mismatch}
\end{align}

The mismatch is related to the signal-to-noise ratio of the difference between the two signals $\rho_\mathrm{diff}$ through~\cite{McWilliams2010}
\begin{align}
    \rho^2_\mathrm{diff} &= 2 \rho^2_\mathrm{sig} \mathcal{M}, \label{eqn:rhodiff}
\end{align}
where $\rho_\mathrm{sig}$ is the signal-to-noise ratio of a detected signal (not to be confused with the matter density). A widely used criterion for the two signals to be distinguishable is $\rho_\mathrm{diff}\geq 1$ \cite{Lindblom2008,Prrer2020}. It should be noted that this does not relate to the ability to observe a signal amongst some noise, nor to using a given signal to estimate the specific model differences. Instead it speaks to whether or not one is able distinguish between one model and another, knowing that there is a signal present in the data.

For our purposes, we want to know what signal-to-noise ratio is required for us to be able to distinguish between the two reaction limits, so we rearrange \cref{eqn:rhodiff} to provide the condition 
\begin{align}
    \rho_\mathrm{sig} &\geq \frac{1}{\sqrt{2 \mathcal{M}}}. \label{eqn:rhosig}
\end{align}
To reiterate, if this condition is satisfied for a given detected signal we should be able to differentiate between fast- and slow-reaction limit behaviour. We define $\rho_\mathrm{req}=1/\sqrt{2\mathcal{M}}$ as the smallest signal-to-noise ratio required to satisfy this condition.

\section{Results \& Discussion}

\begin{figure}[t]
\centering
\includegraphics[width=0.45\textwidth]{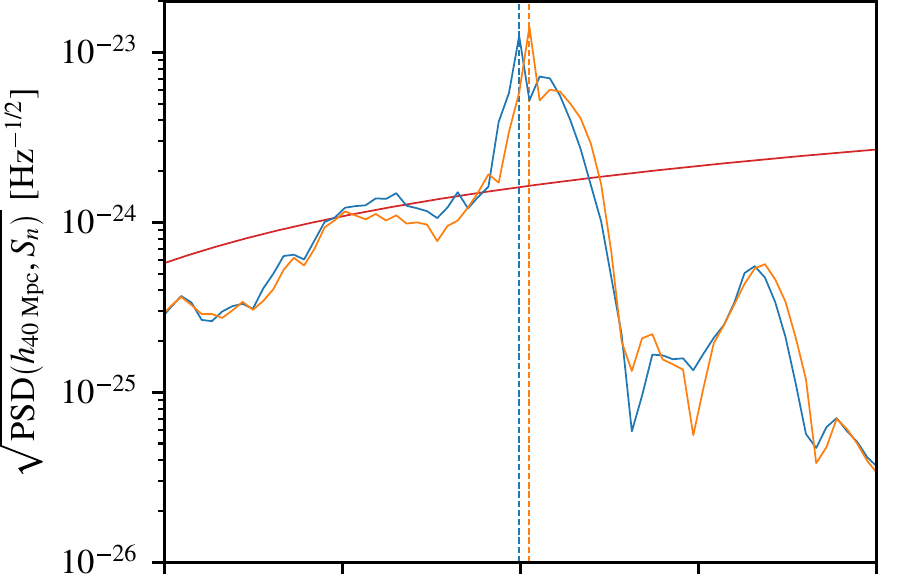}
\includegraphics[width=0.45\textwidth]{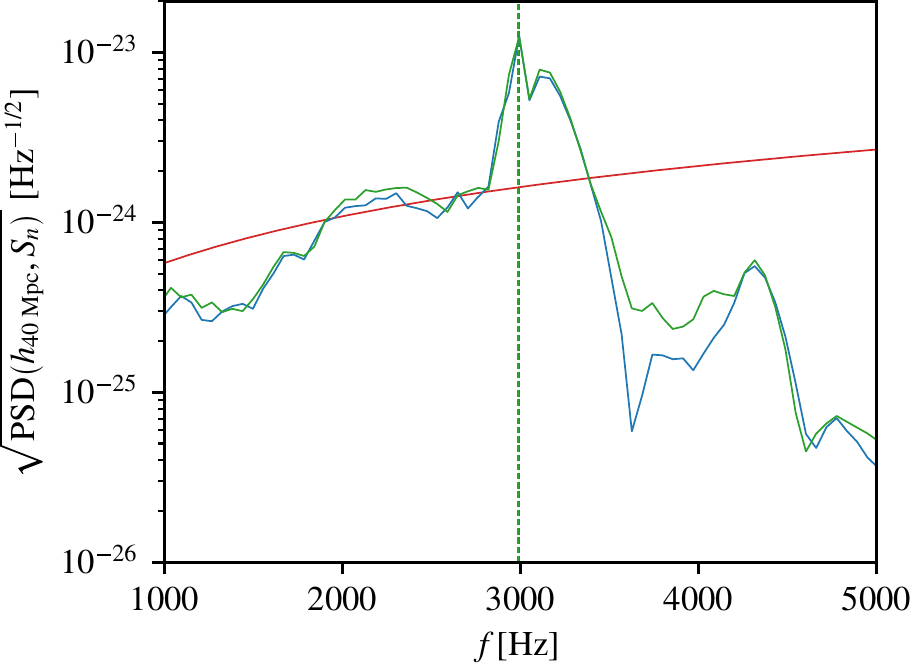}
\caption{Upper panel: Square-root power spectral density plots of the recovered waveforms from slow- and fast-reaction-limit simulations (blue and orange curves, respectively). Lower panel: The same comparison for the high- and low-resolution simulations (blue and green curves, respectively), both in the slow reaction limit. Both panels: Peak frequencies for each curve calculated using the MacLeod method~\cite{Macleod1998} are shown with dashed lines. The ET-D design sensitivity curve for the Einstein Telescope~\cite{Hild2011} is shown in red. The waveforms are normalised to a distance of $40\ \mathrm{Mpc}$ and assume the source and detector are perfectly aligned.}
\label{fig:PSD_Cut}
\end{figure}

\textit{Waveform comparison}.---
In the upper panel of \cref{fig:PSD_Cut} we show the post-merger gravitational wave output of the fast- and slow-reaction-limit simulations, alongside the suggested ET-D noise curve for the Einstein Telescope~\cite{Hild2011} (which we use as the detector noise in all mismatch calculations), assuming a polar aligned detector at a distance of $40\ \mathrm{Mpc}$. While the two signals have similar overall shapes in the frequency domain, we see that the peak frequencies for the two simulations are visibly different: $2992 \pm 8\ \mathrm{Hz}$ in the slow-limit simulation, and $3050 \pm 10\ \mathrm{Hz}$ in the fast-limit simulation, giving a difference of $\Delta f = 58 \pm 13\ \mathrm{Hz}$. Peak frequencies are calculated using the method of MacLeod~\cite{Macleod1998}, and checked using the method of Quinn~\cite{Quinn1997} (in all cases the estimate from the second method is within the error bounds of the first). The computed mismatch between these signals is $\mathcal{M} = 36\%$, which, using \cref{eqn:rhosig}, gives a required signal to noise ratio for the two signals to be distinguishable of $\rho_\mathrm{req} = 1.2$. As this is lower than any reasonable threshold to claim a detection, if the post-merger signal is detected then the effect of the reactions must be taken into account.

As a sanity check, we also performed a simulation with $20\%$ coarser resolution but otherwise identical to the slow limit simulation, the output of which is plotted (alongside the original slow-limit results) in the lower panel of \cref{fig:PSD_Cut}. Comparing these results with the original slow limit simulation we find a mismatch on the order of $1\%$, and the lower resolution simulation has a measured peak frequency of $2992 \pm 6\ \mathrm{Hz}$. The mismatch in this case is much lower than the mismatch between the fast- and slow-limit simulations, and the peak frequency extracted from the low resolution simulation is within the error bounds of the high-resolution simulation, so we are confident that the numerical error due to the finite resolution of the simulations is much less than the difference driven by the contrasting physics models in the two cases.

As a further check we have performed the above analysis using the full gravitational-wave data from each simulation (as opposed to just the post-merger data) after high-pass filtering (the methodology used in~\cite{Takami2015}) and obtained mismatches of $33\%$ and $1\%$ for the fast-/slow-limit and high-/low-resolution combinations, respectively. All of the recovered peak frequencies were within the error bounds of their counterparts given above.

\textit{Effect on the equation of state}.---
Having established that there is a measurable difference between the gravitational wave signals in the two reaction limits, it is natural to question what drives this difference. We start by using a simple model motivated by the results of the simulations to explore what effect being out of equilibrium has on an isolated star. By constructing two 1-D equations of state from the full 3-D table, one in- and one out-of-equilibrium, we can see what effect being out of equilibrium has on the structure of the stars. For the out-of-equilibrium equation of state we solve the composition for a constant $\mu_\Delta = \mu_\mathrm{n} - (\mu_\mathrm{p}+\mu_\mathrm{e})$ throughout the table. This value of the offset need not be wholly representative of the full simulations, but for the sake of comparison we use the simulation data to inform our choice. In figure 5 of \cite{Hammond2021} we see the deviation from chemical equilibrium in the two stars around the time of merger. At $5\ \mathrm{ms}$ post merger most of the core matter is in the $\mu_\Delta = 20-30\ \mathrm{MeV}$ range (excluding the hotspots), hence we choose $\mu_\Delta = 20\ \mathrm{MeV}$ as the deviation from equilibrium for our out-of-equilibrium equation of state. Finally, we obtain a 1-D equation of state by choosing a temperature of $T = 5\ \mathrm{MeV}$.

Using LORENE~\cite{LORENE} we first construct a rotating star using the equilibrium equation of state. We choose the mass of the star such that the central baryon number density is similar to that found in our simulation ($3.5 n_\mathrm{sat}$), giving a total baryon mass for the star of $1.7 \mathrm{M}_\odot$, and an arbitrary rotation frequency of $500\ \mathrm{Hz}$ (significant, but well below the Keplerian frequency). This gives us a total angular momentum. We then construct a rotating star using the out-of-equilibrium equation of state conserving the total baryon mass and angular momentum from the equilibrium star, giving us a different rotation frequency. 

We find that to conserve angular momentum, the rotation frequency must be reduced to $496\ \mathrm{Hz}$ (a fractional difference of $\sim 1\%$). We repeated this procedure for the DD2~\cite{Hempel2010,Typel2010}, SFHx~\cite{Steiner2013}, and SLy4~\cite{Schneider2017,Chabanat1998} equations of state, and obtained similar results, with the change in rotation frequency required being of order $1\%$, and always in the same direction: out-of-equilibrium stars need a lower frequency to match the angular momentum of the respective in-equilibrium stars due to an increase in moment of inertia. This increase is caused by a softening of the equation of state at core densities as the matter is taken out of equilibrium, creating a flatter density profile, and thus more of the mass of the star is located further from the axis of rotation.

Having observed similar effects in several equations of state, one might ask whether this effect is truly general. We will do this through the adiabatic index
\begin{align}
    \Gamma &= \left.\frac{\partial \ln p}{\partial \ln n_b}\right|_{\mathcal{S},Y_\mathrm{e}} = \frac{n_b}{p} \left.\frac{\partial p}{\partial n_b}\right|_{\mathcal{S},Y_\mathrm{e}}, \label{eqn:gammadef}
\end{align}
where $\mathcal{S}$ is the specific entropy per baryon. Taylor expanding around equilibrium and keeping only the first term we obtain
\begin{align}
    \Gamma \left( \mu_\Delta \right) &= \Gamma \left( 0 \right) + \mu_\Delta \left.\frac{\partial \Gamma\left(0\right)}{\partial \mu_\Delta}\right|_{\mathcal{S},n_b} + \mathcal{O} \left( \mu^2_\Delta \right), \label{eqn:Gamma_taylor}
\end{align}
so the change in stiffness when moving out of equilibrium will depend on the sign of the second term. The sign is dependent on the equation of state used, however assuming a simple Fermi gas model gives a negative value for all densities. Using the more realistic BSk density functionals \cite{Goriely2013} we also obtain a negative sign at the densities of interest.

\textit{Effect on the matter distribution}.---
Having examined some simpler systems to determine a possible source for the frequency difference presented above, we can also look to see whether the softening effect is visible in the full merger simulation. The na\"{i}ve approach would be to measure the moment of inertia of the whole system in each case. However, after merger the moment of inertia is dominated by low density matter far from the remnant, so we need some local measure that we can apply only to the remnant. 

Starting from the Newtonian definition of the moment of inertia for a mass distribution along the $z$-axis
\begin{align}
    I_z (\rho) = \iiint_V \rho(x,y,z) (x^2 + y^2) \mathrm{d}V
\end{align}
we substitute in the proper mass density $\rho W$, and correct the volume element to obtain
\begin{align}
    I^z = \iiint_V \rho W (x^1 x^1 + x^2 x^2) \sqrt{\gamma} \mathrm{d}^3 x. \label{eqn:MoI_3D}
\end{align}

To better see the differences between the remnants we will use a density cutoff to mask the effects of low density matter at large radius, and to determine a relevant density for the cutoff we will look at how the contribution to the moment of inertia varies in the $x$-$y$ plane of the two stars. To do this we calculate the infinitesimal contribution to the total moment of inertia of the loop at radius $r$, $I^\mathrm{loop}(r)$, then integrate radially outwards (to obtain $I^\mathrm{disc}(r)$) to find the density at which the difference between the two simulations settles down. 
We find that the difference between $I^\mathrm{disc}_\mathrm{OoE}$ and $I^\mathrm{disc}_\mathrm{NSE}$ has settled down by $\rho \sim 10^{-2} - 10^{-3} \rho_\mathrm{sat}$.

\begin{figure}[t]
\centering
\includegraphics[width=0.475\textwidth]{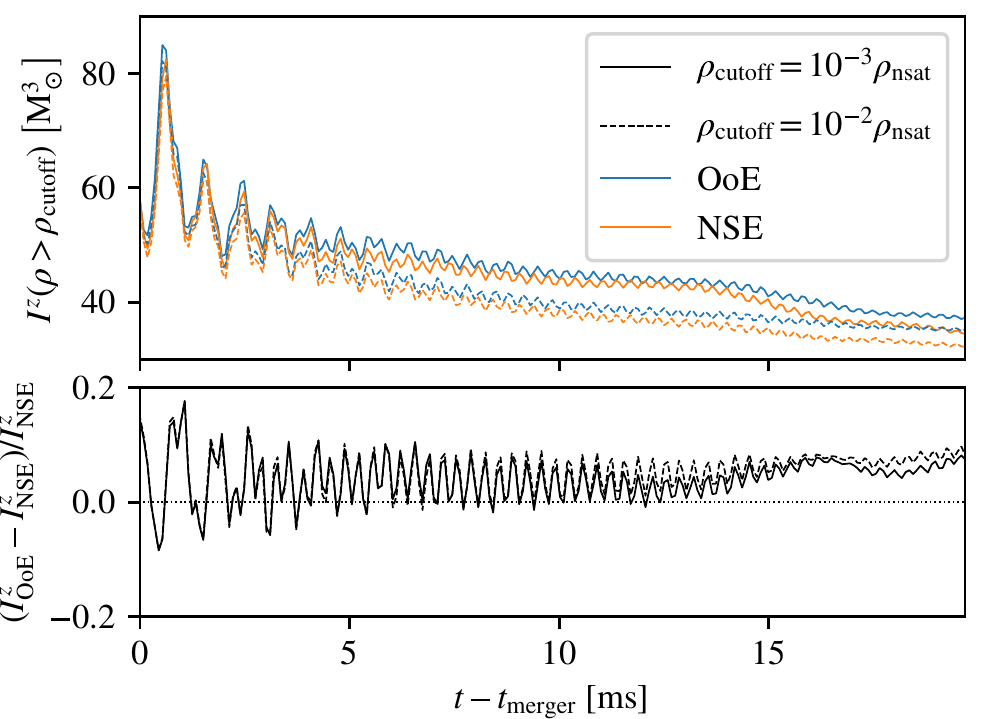}
\caption{Upper panel: Evolution of the total moment of inertia (see \cref{eqn:MoI_3D}) ignoring matter with rest mass density below $\rho_\mathrm{cutoff}$ for slow- and fast-reaction limit simulations (labelled $\mathrm{OoE}$ and $\mathrm{NSE}$, solid and dotted lines, respectively). Lower panel: Relative difference between results for slow- and fast-limit simulations. }
\label{fig:MoI_3D}
\end{figure}

Ignoring densities below this cutoff, we can then apply \cref{eqn:MoI_3D} to the whole domain, the results of which we plot in \cref{fig:MoI_3D}. This shows that when considering only the densities above $\rho_\mathrm{cutoff} \sim 10^{-2} - 10^{-3} \rho_\mathrm{sat}$, the moment of inertia of the remnant is a few percent higher in the out-of-equilibrium simulation, which would intuitively lead to a few percent decrease in the frequency at which the remnant rotates.

As we have shown above, this increase in moment of inertia can be driven by a softening of the equation of state. We have also shown that under an adiabatic compression (with no reactions) this softening is expected, independent of the equation of state. This is robust evidence that the phase of the gravitational wave-signal is sufficiently sensitive to models of weak reactions that these must be accounted for in signal templates for the next generation gravitational-wave observatories, like the Einstein Telescope and the Cosmic Explorer.

\begin{acknowledgments}
NA and IH  are  grateful for support from STFC via grant numbers ST/R00045X/1 and ST/V000551/1.

This letter has been given ET TDS number ET-0099A-22.

Source files for the modified version of the Einstein Toolkit used in this work can be found at \href{https://doi.org/10.5281/zenodo.5469497}{DOI: 10.5281/zenodo.5469497}. Parfiles and plotting scripts for the simulations discussed and figures presented in this work can be found at \href{https://doi.org/10.5281/zenodo.6532867}{DOI: 10.5281/zenodo.6532867}.
\end{acknowledgments}

\bibliography{bib}

\begin{thebibliography}{41}
\expandafter\ifx\csname natexlab\endcsname\relax\def\natexlab#1{#1}\fi
\expandafter\ifx\csname bibnamefont\endcsname\relax
  \def\bibnamefont#1{#1}\fi
\expandafter\ifx\csname bibfnamefont\endcsname\relax
  \def\bibfnamefont#1{#1}\fi
\expandafter\ifx\csname citenamefont\endcsname\relax
  \def\citenamefont#1{#1}\fi
\expandafter\ifx\csname url\endcsname\relax
  \def\url#1{\texttt{#1}}\fi
\expandafter\ifx\csname urlprefix\endcsname\relax\def\urlprefix{URL }\fi
\providecommand{\bibinfo}[2]{#2}
\providecommand{\eprint}[2][]{\url{#2}}

\bibitem[{\citenamefont{Abbott et~al.}(2017)}]{Abbott2017}
\bibinfo{author}{\bibfnamefont{B.~P.} \bibnamefont{Abbott}}
  \bibnamefont{et~al.}, \bibinfo{journal}{Physical Review Letters}
  \textbf{\bibinfo{volume}{119}} (\bibinfo{year}{2017}).

\bibitem[{\citenamefont{Baiotti and Rezzolla}(2017)}]{Baiotti2017}
\bibinfo{author}{\bibfnamefont{L.}~\bibnamefont{Baiotti}} \bibnamefont{and}
  \bibinfo{author}{\bibfnamefont{L.}~\bibnamefont{Rezzolla}},
  \bibinfo{journal}{Reports on Progress in Physics}
  \textbf{\bibinfo{volume}{80}}, \bibinfo{pages}{096901}
  (\bibinfo{year}{2017}).

\bibitem[{\citenamefont{Bernuzzi}(2020)}]{Bernuzzi2020}
\bibinfo{author}{\bibfnamefont{S.}~\bibnamefont{Bernuzzi}},
  \bibinfo{journal}{General Relativity and Gravitation}
  \textbf{\bibinfo{volume}{52}} (\bibinfo{year}{2020}).

\bibitem[{\citenamefont{Punturo et~al.}(2010)}]{Punturo2010}
\bibinfo{author}{\bibfnamefont{M.}~\bibnamefont{Punturo}} \bibnamefont{et~al.},
  \bibinfo{journal}{Classical and Quantum Gravity}
  \textbf{\bibinfo{volume}{27}}, \bibinfo{pages}{194002}
  (\bibinfo{year}{2010}).

\bibitem[{\citenamefont{Hild et~al.}(2011)}]{Hild2011}
\bibinfo{author}{\bibfnamefont{S.}~\bibnamefont{Hild}} \bibnamefont{et~al.},
  \bibinfo{journal}{Classical and Quantum Gravity}
  \textbf{\bibinfo{volume}{28}} (\bibinfo{year}{2011}).

\bibitem[{ET()}]{ET}
\emph{\bibinfo{title}{The {E}instein {T}elescope}},
  \urlprefix\url{http://www.et-gw.eu/}.

\bibitem[{\citenamefont{Reitze et~al.}(2019)}]{Reitze2019}
\bibinfo{author}{\bibfnamefont{D.}~\bibnamefont{Reitze}} \bibnamefont{et~al.},
  \emph{\bibinfo{title}{Cosmic explorer: The u.s. contribution to
  gravitational-wave astronomy beyond ligo}} (\bibinfo{year}{2019}),
  \eprint{arXiv:1907.04833}.

\bibitem[{CE()}]{CE}
\emph{\bibinfo{title}{{C}osmic {E}xplorer}},
  \urlprefix\url{https://cosmicexplorer.org/}.

\bibitem[{\citenamefont{Baiotti}(2022)}]{Baiotti2022}
\bibinfo{author}{\bibfnamefont{L.}~\bibnamefont{Baiotti}},
  \bibinfo{journal}{Arabian Journal of Mathematics}  (\bibinfo{year}{2022}).

\bibitem[{\citenamefont{Dietrich et~al.}(2018)\citenamefont{Dietrich, Radice,
  Bernuzzi, Zappa, Perego, Br\"{u}gmann, Chaurasia, Dudi, Tichy, and
  Ujevic}}]{Dietrich2018}
\bibinfo{author}{\bibfnamefont{T.}~\bibnamefont{Dietrich}},
  \bibinfo{author}{\bibfnamefont{D.}~\bibnamefont{Radice}},
  \bibinfo{author}{\bibfnamefont{S.}~\bibnamefont{Bernuzzi}},
  \bibinfo{author}{\bibfnamefont{F.}~\bibnamefont{Zappa}},
  \bibinfo{author}{\bibfnamefont{A.}~\bibnamefont{Perego}},
  \bibinfo{author}{\bibfnamefont{B.}~\bibnamefont{Br\"{u}gmann}},
  \bibinfo{author}{\bibfnamefont{S.~V.} \bibnamefont{Chaurasia}},
  \bibinfo{author}{\bibfnamefont{R.}~\bibnamefont{Dudi}},
  \bibinfo{author}{\bibfnamefont{W.}~\bibnamefont{Tichy}}, \bibnamefont{and}
  \bibinfo{author}{\bibfnamefont{M.}~\bibnamefont{Ujevic}},
  \bibinfo{journal}{Classical and Quantum Gravity}
  \textbf{\bibinfo{volume}{35}}, \bibinfo{pages}{24LT01}
  (\bibinfo{year}{2018}).

\bibitem[{\citenamefont{Cabez{\'{o}}n et~al.}(2018)\citenamefont{Cabez{\'{o}}n,
  Pan, Liebend\"{o}rfer, Kuroda, Ebinger, Heinimann, Perego, and
  Thielemann}}]{Cabezn2018}
\bibinfo{author}{\bibfnamefont{R.~M.} \bibnamefont{Cabez{\'{o}}n}},
  \bibinfo{author}{\bibfnamefont{K.-C.} \bibnamefont{Pan}},
  \bibinfo{author}{\bibfnamefont{M.}~\bibnamefont{Liebend\"{o}rfer}},
  \bibinfo{author}{\bibfnamefont{T.}~\bibnamefont{Kuroda}},
  \bibinfo{author}{\bibfnamefont{K.}~\bibnamefont{Ebinger}},
  \bibinfo{author}{\bibfnamefont{O.}~\bibnamefont{Heinimann}},
  \bibinfo{author}{\bibfnamefont{A.}~\bibnamefont{Perego}}, \bibnamefont{and}
  \bibinfo{author}{\bibfnamefont{F.-K.} \bibnamefont{Thielemann}},
  \bibinfo{journal}{Astronomy \& Astrophysics} \textbf{\bibinfo{volume}{619}},
  \bibinfo{pages}{A118} (\bibinfo{year}{2018}).

\bibitem[{\citenamefont{Pan et~al.}(2018)\citenamefont{Pan, Mattes, O'Connor,
  Couch, Perego, and Arcones}}]{Pan2018}
\bibinfo{author}{\bibfnamefont{K.-C.} \bibnamefont{Pan}},
  \bibinfo{author}{\bibfnamefont{C.}~\bibnamefont{Mattes}},
  \bibinfo{author}{\bibfnamefont{E.~P.} \bibnamefont{O'Connor}},
  \bibinfo{author}{\bibfnamefont{S.~M.} \bibnamefont{Couch}},
  \bibinfo{author}{\bibfnamefont{A.}~\bibnamefont{Perego}}, \bibnamefont{and}
  \bibinfo{author}{\bibfnamefont{A.}~\bibnamefont{Arcones}},
  \bibinfo{journal}{Journal of Physics G: Nuclear and Particle Physics}
  \textbf{\bibinfo{volume}{46}}, \bibinfo{pages}{014001}
  (\bibinfo{year}{2018}).

\bibitem[{\citenamefont{Richers et~al.}(2017)\citenamefont{Richers, Nagakura,
  Ott, Dolence, Sumiyoshi, and Yamada}}]{Richers2017}
\bibinfo{author}{\bibfnamefont{S.}~\bibnamefont{Richers}},
  \bibinfo{author}{\bibfnamefont{H.}~\bibnamefont{Nagakura}},
  \bibinfo{author}{\bibfnamefont{C.~D.} \bibnamefont{Ott}},
  \bibinfo{author}{\bibfnamefont{J.}~\bibnamefont{Dolence}},
  \bibinfo{author}{\bibfnamefont{K.}~\bibnamefont{Sumiyoshi}},
  \bibnamefont{and} \bibinfo{author}{\bibfnamefont{S.}~\bibnamefont{Yamada}},
  \bibinfo{journal}{The Astrophysical Journal} \textbf{\bibinfo{volume}{847}},
  \bibinfo{pages}{133} (\bibinfo{year}{2017}).

\bibitem[{\citenamefont{Cusinato et~al.}(2021)\citenamefont{Cusinato,
  Guercilena, Perego, Logoteta, Radice, Bernuzzi, and Ansoldi}}]{Cusinato2021}
\bibinfo{author}{\bibfnamefont{M.}~\bibnamefont{Cusinato}},
  \bibinfo{author}{\bibfnamefont{F.~M.} \bibnamefont{Guercilena}},
  \bibinfo{author}{\bibfnamefont{A.}~\bibnamefont{Perego}},
  \bibinfo{author}{\bibfnamefont{D.}~\bibnamefont{Logoteta}},
  \bibinfo{author}{\bibfnamefont{D.}~\bibnamefont{Radice}},
  \bibinfo{author}{\bibfnamefont{S.}~\bibnamefont{Bernuzzi}}, \bibnamefont{and}
  \bibinfo{author}{\bibfnamefont{S.}~\bibnamefont{Ansoldi}},
  \emph{\bibinfo{title}{Neutrino emission from binary neutron star mergers:
  characterizing light curves and mean energies}} (\bibinfo{year}{2021}),
  \eprint{arXiv:2111.13005}.

\bibitem[{\citenamefont{Mezzacappa et~al.}(2020)\citenamefont{Mezzacappa,
  Endeve, Messer, and Bruenn}}]{Mezzacappa2020}
\bibinfo{author}{\bibfnamefont{A.}~\bibnamefont{Mezzacappa}},
  \bibinfo{author}{\bibfnamefont{E.}~\bibnamefont{Endeve}},
  \bibinfo{author}{\bibfnamefont{O.~E.~B.} \bibnamefont{Messer}},
  \bibnamefont{and} \bibinfo{author}{\bibfnamefont{S.~W.}
  \bibnamefont{Bruenn}}, \bibinfo{journal}{Living Reviews in Computational
  Astrophysics} \textbf{\bibinfo{volume}{6}} (\bibinfo{year}{2020}).

\bibitem[{\citenamefont{Celora et~al.}(2022)\citenamefont{Celora, Hawke,
  Hammond, Andersson, and Comer}}]{Celora2022}
\bibinfo{author}{\bibfnamefont{T.}~\bibnamefont{Celora}},
  \bibinfo{author}{\bibfnamefont{I.}~\bibnamefont{Hawke}},
  \bibinfo{author}{\bibfnamefont{P.~C.} \bibnamefont{Hammond}},
  \bibinfo{author}{\bibfnamefont{N.}~\bibnamefont{Andersson}},
  \bibnamefont{and} \bibinfo{author}{\bibfnamefont{G.~L.} \bibnamefont{Comer}},
  \bibinfo{journal}{Physical Review D} \textbf{\bibinfo{volume}{105}}
  (\bibinfo{year}{2022}).

\bibitem[{\citenamefont{Radice et~al.}(2022)\citenamefont{Radice, Bernuzzi,
  Perego, and Haas}}]{Radice2022}
\bibinfo{author}{\bibfnamefont{D.}~\bibnamefont{Radice}},
  \bibinfo{author}{\bibfnamefont{S.}~\bibnamefont{Bernuzzi}},
  \bibinfo{author}{\bibfnamefont{A.}~\bibnamefont{Perego}}, \bibnamefont{and}
  \bibinfo{author}{\bibfnamefont{R.}~\bibnamefont{Haas}},
  \bibinfo{journal}{Monthly Notices of the Royal Astronomical Society}
  \textbf{\bibinfo{volume}{512}}, \bibinfo{pages}{1499} (\bibinfo{year}{2022}).

\bibitem[{\citenamefont{Camilletti et~al.}(2022)\citenamefont{Camilletti,
  Chiesa, Ricigliano, Perego, Lippold, Padamata, Bernuzzi, Radice, Logoteta,
  and Guercilena}}]{Camilletti2022}
\bibinfo{author}{\bibfnamefont{A.}~\bibnamefont{Camilletti}},
  \bibinfo{author}{\bibfnamefont{L.}~\bibnamefont{Chiesa}},
  \bibinfo{author}{\bibfnamefont{G.}~\bibnamefont{Ricigliano}},
  \bibinfo{author}{\bibfnamefont{A.}~\bibnamefont{Perego}},
  \bibinfo{author}{\bibfnamefont{L.~C.} \bibnamefont{Lippold}},
  \bibinfo{author}{\bibfnamefont{S.}~\bibnamefont{Padamata}},
  \bibinfo{author}{\bibfnamefont{S.}~\bibnamefont{Bernuzzi}},
  \bibinfo{author}{\bibfnamefont{D.}~\bibnamefont{Radice}},
  \bibinfo{author}{\bibfnamefont{D.}~\bibnamefont{Logoteta}}, \bibnamefont{and}
  \bibinfo{author}{\bibfnamefont{F.~M.} \bibnamefont{Guercilena}},
  \emph{\bibinfo{title}{Numerical relativity simulations of the neutron star
  merger gw190425: microphysics and mass ratio effects}}
  (\bibinfo{year}{2022}), \eprint{arXiv:2204.05336}.

\bibitem[{\citenamefont{Palenzuela et~al.}(2022)\citenamefont{Palenzuela,
  Liebling, and Miñano}}]{Palenzuela2022}
\bibinfo{author}{\bibfnamefont{C.}~\bibnamefont{Palenzuela}},
  \bibinfo{author}{\bibfnamefont{S.~L.} \bibnamefont{Liebling}},
  \bibnamefont{and} \bibinfo{author}{\bibfnamefont{B.}~\bibnamefont{Miñano}},
  \emph{\bibinfo{title}{Large eddy simulations of magnetized mergers of neutron
  stars with neutrinos}} (\bibinfo{year}{2022}), \eprint{arXiv:2204.02721}.

\bibitem[{\citenamefont{Hayashi et~al.}(2021)\citenamefont{Hayashi,
  Fujibayashi, Kiuchi, Kyutoku, Sekiguchi, and Shibata}}]{Hayashi2021}
\bibinfo{author}{\bibfnamefont{K.}~\bibnamefont{Hayashi}},
  \bibinfo{author}{\bibfnamefont{S.}~\bibnamefont{Fujibayashi}},
  \bibinfo{author}{\bibfnamefont{K.}~\bibnamefont{Kiuchi}},
  \bibinfo{author}{\bibfnamefont{K.}~\bibnamefont{Kyutoku}},
  \bibinfo{author}{\bibfnamefont{Y.}~\bibnamefont{Sekiguchi}},
  \bibnamefont{and} \bibinfo{author}{\bibfnamefont{M.}~\bibnamefont{Shibata}},
  \emph{\bibinfo{title}{General-relativistic neutrino-radiation
  magnetohydrodynamics simulation of black hole-neutron star mergers for
  seconds}} (\bibinfo{year}{2021}), \eprint{arXiv:2111.04621}.

\bibitem[{\citenamefont{Banyuls et~al.}(1997)\citenamefont{Banyuls, Font,
  Ibanez, Marti, and Miralles}}]{Banyuls1997}
\bibinfo{author}{\bibfnamefont{F.}~\bibnamefont{Banyuls}},
  \bibinfo{author}{\bibfnamefont{J.~A.} \bibnamefont{Font}},
  \bibinfo{author}{\bibfnamefont{J.~M.} \bibnamefont{Ibanez}},
  \bibinfo{author}{\bibfnamefont{J.~M.} \bibnamefont{Marti}}, \bibnamefont{and}
  \bibinfo{author}{\bibfnamefont{J.~A.} \bibnamefont{Miralles}},
  \bibinfo{journal}{Ap. J.} \textbf{\bibinfo{volume}{476}},
  \bibinfo{pages}{221} (\bibinfo{year}{1997}).

\bibitem[{\citenamefont{Baiotti et~al.}(2005)\citenamefont{Baiotti, Hawke,
  Montero, L{\"o}ffler, Rezzolla, Stergioulas, Font, and Seidel}}]{Baiotti2005}
\bibinfo{author}{\bibfnamefont{L.}~\bibnamefont{Baiotti}},
  \bibinfo{author}{\bibfnamefont{I.}~\bibnamefont{Hawke}},
  \bibinfo{author}{\bibfnamefont{P.~J.} \bibnamefont{Montero}},
  \bibinfo{author}{\bibfnamefont{F.}~\bibnamefont{L{\"o}ffler}},
  \bibinfo{author}{\bibfnamefont{L.}~\bibnamefont{Rezzolla}},
  \bibinfo{author}{\bibfnamefont{N.}~\bibnamefont{Stergioulas}},
  \bibinfo{author}{\bibfnamefont{J.~A.} \bibnamefont{Font}}, \bibnamefont{and}
  \bibinfo{author}{\bibfnamefont{E.}~\bibnamefont{Seidel}},
  \bibinfo{journal}{Phys. Rev. D} \textbf{\bibinfo{volume}{71}},
  \bibinfo{pages}{024035} (\bibinfo{year}{2005}).

\bibitem[{\citenamefont{Alford and Harris}(2018)}]{Alford2018a}
\bibinfo{author}{\bibfnamefont{M.~G.} \bibnamefont{Alford}} \bibnamefont{and}
  \bibinfo{author}{\bibfnamefont{S.~P.} \bibnamefont{Harris}},
  \bibinfo{journal}{Phys. Rev. C} \textbf{\bibinfo{volume}{98}},
  \bibinfo{pages}{065806} (\bibinfo{year}{2018}).

\bibitem[{\citenamefont{Hammond et~al.}(2021)\citenamefont{Hammond, Hawke, and
  Andersson}}]{Hammond2021}
\bibinfo{author}{\bibfnamefont{P.}~\bibnamefont{Hammond}},
  \bibinfo{author}{\bibfnamefont{I.}~\bibnamefont{Hawke}}, \bibnamefont{and}
  \bibinfo{author}{\bibfnamefont{N.}~\bibnamefont{Andersson}},
  \bibinfo{journal}{Phys. Rev. D} \textbf{\bibinfo{volume}{104}},
  \bibinfo{pages}{103006} (\bibinfo{year}{2021}).

\bibitem[{\citenamefont{Etienne et~al.}(accessed 2021)}]{EinsteinToolkit}
\bibinfo{author}{\bibfnamefont{Z.}~\bibnamefont{Etienne}} \bibnamefont{et~al.},
  \emph{\bibinfo{title}{{The Einstein Toolkit}}} (\bibinfo{year}{accessed
  2021}), \urlprefix\url{http://einsteintoolkit.org/index.html}.

\bibitem[{\citenamefont{Gourgoulhon et~al.}(accessed
  2022)\citenamefont{Gourgoulhon, Grandcl\'{e}ment, and Novak}}]{LORENE}
\bibinfo{author}{\bibfnamefont{E.}~\bibnamefont{Gourgoulhon}},
  \bibinfo{author}{\bibfnamefont{P.}~\bibnamefont{Grandcl\'{e}ment}},
  \bibnamefont{and} \bibinfo{author}{\bibfnamefont{J.}~\bibnamefont{Novak}},
  \emph{\bibinfo{title}{{LORENE}}} (\bibinfo{year}{accessed 2022}),
  \urlprefix\url{https://lorene.obspm.fr/}.

\bibitem[{\citenamefont{Reisswig and Pollney}(2011)}]{Reisswig2011}
\bibinfo{author}{\bibfnamefont{C.}~\bibnamefont{Reisswig}} \bibnamefont{and}
  \bibinfo{author}{\bibfnamefont{D.}~\bibnamefont{Pollney}},
  \bibinfo{journal}{Classical and Quantum Gravity}
  \textbf{\bibinfo{volume}{28}} (\bibinfo{year}{2011}).

\bibitem[{\citenamefont{Lindblom et~al.}(2008)\citenamefont{Lindblom, Owen, and
  Brown}}]{Lindblom2008}
\bibinfo{author}{\bibfnamefont{L.}~\bibnamefont{Lindblom}},
  \bibinfo{author}{\bibfnamefont{B.~J.} \bibnamefont{Owen}}, \bibnamefont{and}
  \bibinfo{author}{\bibfnamefont{D.~A.} \bibnamefont{Brown}},
  \bibinfo{journal}{Phys. Rev. D} \textbf{\bibinfo{volume}{78}},
  \bibinfo{pages}{124020} (\bibinfo{year}{2008}).

\bibitem[{\citenamefont{McWilliams et~al.}(2010)\citenamefont{McWilliams,
  Kelly, and Baker}}]{McWilliams2010}
\bibinfo{author}{\bibfnamefont{S.~T.} \bibnamefont{McWilliams}},
  \bibinfo{author}{\bibfnamefont{B.~J.} \bibnamefont{Kelly}}, \bibnamefont{and}
  \bibinfo{author}{\bibfnamefont{J.~G.} \bibnamefont{Baker}},
  \bibinfo{journal}{Phys. Rev. D} \textbf{\bibinfo{volume}{82}},
  \bibinfo{pages}{024014} (\bibinfo{year}{2010}).

\bibitem[{\citenamefont{Baird et~al.}(2013)}]{Baird2013}
\bibinfo{author}{\bibfnamefont{E.}~\bibnamefont{Baird}} \bibnamefont{et~al.},
  \bibinfo{journal}{Phys. Rev. D} \textbf{\bibinfo{volume}{87}},
  \bibinfo{pages}{024035} (\bibinfo{year}{2013}).

\bibitem[{\citenamefont{Kumar et~al.}(2015)}]{Kumar2015}
\bibinfo{author}{\bibfnamefont{P.}~\bibnamefont{Kumar}} \bibnamefont{et~al.},
  \bibinfo{journal}{Phys. Rev. D} \textbf{\bibinfo{volume}{92}},
  \bibinfo{pages}{102001} (\bibinfo{year}{2015}).

\bibitem[{\citenamefont{P\"{u}rrer and Haster}(2020)}]{Prrer2020}
\bibinfo{author}{\bibfnamefont{M.}~\bibnamefont{P\"{u}rrer}} \bibnamefont{and}
  \bibinfo{author}{\bibfnamefont{C.-J.} \bibnamefont{Haster}},
  \bibinfo{journal}{Physical Review Research} \textbf{\bibinfo{volume}{2}}
  (\bibinfo{year}{2020}).

\bibitem[{\citenamefont{Macleod}(1998)}]{Macleod1998}
\bibinfo{author}{\bibfnamefont{M.}~\bibnamefont{Macleod}},
  \bibinfo{journal}{IEEE Transactions on Signal Processing}
  \textbf{\bibinfo{volume}{46}}, \bibinfo{pages}{141} (\bibinfo{year}{1998}).

\bibitem[{\citenamefont{Quinn}(1997)}]{Quinn1997}
\bibinfo{author}{\bibfnamefont{B.}~\bibnamefont{Quinn}}, \bibinfo{journal}{IEEE
  Transactions on Signal Processing} \textbf{\bibinfo{volume}{45}},
  \bibinfo{pages}{814} (\bibinfo{year}{1997}).

\bibitem[{\citenamefont{Takami et~al.}(2015)\citenamefont{Takami, Rezzolla, and
  Baiotti}}]{Takami2015}
\bibinfo{author}{\bibfnamefont{K.}~\bibnamefont{Takami}},
  \bibinfo{author}{\bibfnamefont{L.}~\bibnamefont{Rezzolla}}, \bibnamefont{and}
  \bibinfo{author}{\bibfnamefont{L.}~\bibnamefont{Baiotti}},
  \bibinfo{journal}{Phys. Rev. D} \textbf{\bibinfo{volume}{91}},
  \bibinfo{pages}{064001} (\bibinfo{year}{2015}).

\bibitem[{\citenamefont{Hempel and Schaffner-Bielich}(2010)}]{Hempel2010}
\bibinfo{author}{\bibfnamefont{M.}~\bibnamefont{Hempel}} \bibnamefont{and}
  \bibinfo{author}{\bibfnamefont{J.}~\bibnamefont{Schaffner-Bielich}},
  \bibinfo{journal}{Nuclear Physics A} \textbf{\bibinfo{volume}{837}},
  \bibinfo{pages}{210} (\bibinfo{year}{2010}).

\bibitem[{\citenamefont{Typel et~al.}(2010)\citenamefont{Typel, R\"opke,
  Kl\"ahn, Blaschke, and Wolter}}]{Typel2010}
\bibinfo{author}{\bibfnamefont{S.}~\bibnamefont{Typel}},
  \bibinfo{author}{\bibfnamefont{G.}~\bibnamefont{R\"opke}},
  \bibinfo{author}{\bibfnamefont{T.}~\bibnamefont{Kl\"ahn}},
  \bibinfo{author}{\bibfnamefont{D.}~\bibnamefont{Blaschke}}, \bibnamefont{and}
  \bibinfo{author}{\bibfnamefont{H.~H.} \bibnamefont{Wolter}},
  \bibinfo{journal}{Phys. Rev. C} \textbf{\bibinfo{volume}{81}},
  \bibinfo{pages}{015803} (\bibinfo{year}{2010}).

\bibitem[{\citenamefont{Steiner et~al.}(2013)\citenamefont{Steiner, Hempel, and
  Fischer}}]{Steiner2013}
\bibinfo{author}{\bibfnamefont{A.~W.} \bibnamefont{Steiner}},
  \bibinfo{author}{\bibfnamefont{M.}~\bibnamefont{Hempel}}, \bibnamefont{and}
  \bibinfo{author}{\bibfnamefont{T.}~\bibnamefont{Fischer}},
  \bibinfo{journal}{The Astrophysical Journal} \textbf{\bibinfo{volume}{774}}
  (\bibinfo{year}{2013}).

\bibitem[{\citenamefont{Schneider et~al.}(2017)\citenamefont{Schneider,
  Roberts, and Ott}}]{Schneider2017}
\bibinfo{author}{\bibfnamefont{A.~S.} \bibnamefont{Schneider}},
  \bibinfo{author}{\bibfnamefont{L.~F.} \bibnamefont{Roberts}},
  \bibnamefont{and} \bibinfo{author}{\bibfnamefont{C.~D.} \bibnamefont{Ott}},
  \bibinfo{journal}{Phys. Rev. C} \textbf{\bibinfo{volume}{96}},
  \bibinfo{pages}{065802} (\bibinfo{year}{2017}).

\bibitem[{\citenamefont{Chabanat et~al.}(1998)\citenamefont{Chabanat, Bonche,
  Haensel, Meyer, and Schaeffer}}]{Chabanat1998}
\bibinfo{author}{\bibfnamefont{E.}~\bibnamefont{Chabanat}},
  \bibinfo{author}{\bibfnamefont{P.}~\bibnamefont{Bonche}},
  \bibinfo{author}{\bibfnamefont{P.}~\bibnamefont{Haensel}},
  \bibinfo{author}{\bibfnamefont{J.}~\bibnamefont{Meyer}}, \bibnamefont{and}
  \bibinfo{author}{\bibfnamefont{R.}~\bibnamefont{Schaeffer}},
  \bibinfo{journal}{Nuclear Physics A} \textbf{\bibinfo{volume}{635}},
  \bibinfo{pages}{231} (\bibinfo{year}{1998}).

\bibitem[{\citenamefont{Goriely et~al.}(2013)\citenamefont{Goriely, Chamel, and
  Pearson}}]{Goriely2013}
\bibinfo{author}{\bibfnamefont{S.}~\bibnamefont{Goriely}},
  \bibinfo{author}{\bibfnamefont{N.}~\bibnamefont{Chamel}}, \bibnamefont{and}
  \bibinfo{author}{\bibfnamefont{J.~M.} \bibnamefont{Pearson}},
  \bibinfo{journal}{Phys. Rev. C} \textbf{\bibinfo{volume}{88}},
  \bibinfo{pages}{024308} (\bibinfo{year}{2013}).

\end{thebibliography}

\end{document}